\documentclass[conference,onecolumn,draftclsnofoot]{IEEEtran}

\IEEEoverridecommandlockouts
\usepackage{cite}
\usepackage{amsmath,amssymb,amsfonts}
\usepackage{algorithmic}
\usepackage{graphicx}
\usepackage{textcomp}
\usepackage{xcolor}
\usepackage{lccaps}
\usepackage{optidef}
\usepackage{float}
\usepackage{bm}
\usepackage{subfigure}
\def\BibTeX{{\rm B\kern-.05em{\sc i\kern-.025em b}\kern-.08em
    T\kern-.1667em\lower.7ex\hbox{E}\kern-.125emX}}
\begin{document}

\title{Optimal dispatch schedule for a fast EV charging station with account to supplementary battery health degradation
}
\author{\IEEEauthorblockN{Yihao Wan, Daniel Gebbran, Tomislav Dragičević \\ Department of Wind and Energy Systems, Technical University of Denmark}
}


\maketitle

\begin{abstract}
This paper investigates the usage of battery storage systems in a fast charging station (FCS) for participation in energy markets and charging electrical vehicles (EVs) simultaneously. In particular, we focus on optimizing the scheduling strategies to reduce the overall operational cost of the system over its lifetime by combining the model of battery degradation and energy arbitrage. We implement the battery degradation as a penalty term within an energy arbitrage model and show that the battery degradation plays an important role in the optimal energy dispatch scheduling of the FCS system. In this case study, with different penalty coefficients for the battery degradation penalty term, it is found that including the penalty of battery usage in the scheduling model will reduce the number of small charging/discharging cycles, thereby prolonging the battery lifetime, while maintaining near optimal revenue from grid services.
\end{abstract}


\section{Introduction}
With the increasing electrification of transportation for reducing Greenhouse Gas (GHG) emissions, the electric vehicle (EV) market is taking off \cite{outlook2021accelerating}. Although EVs are becoming more popular, one of the major bottlenecks for the large-scale replacement of Internal Combustion Engine (ICE) is the lack of fast charging infrastructures, particularly on the highways in between cities and rural districts. The fast charging stations are important infrastructures, especially for long-range driving. They can be installed along the highways, providing a refueling experience of EVs similar to the experience of refueling ICE vehicles\cite{morrissey2016future}. However, the load profile of the FCSs is impulsive in nature, which causes adverse impacts on the stability of the grid\cite{mahfouz2019grid}. Moreover, the FCSs are often installed in rural areas with weak grid connections, which requires upgrading electrical grid, resulting in higher installation costs. 

To mitigate the impact of FCS on the grid, battery energy storage systems (BESSs) can be employed\cite{mahfouz2019grid,sbordone2015ev}. BESSs play an increasingly important role in FCSs due to its participation in the followings: (i) energy price arbitrage, (ii) energy demand and supply,  (iii) power supply and demand, (iv) other ancillary services, such as peak shaving, frequency regulation, etc. In particular, a combination of (ii) and (iii) for energy demand and energy supply sides can ensure power balance even under fluctuating load\cite{han2018economic, ding2015value}. In \cite{chen2018coordinated}, a coordinated charging and discharging strategy for a fast charging station is proposed to optimize the economic benefits while the usage of battery is not considered in the scheduling model. Due to the fact that batteries have limited number of charge and discharge cycles, batteries degrade accumulatively during the operations, which should be accounted for in the operational planning problems. The battery degradation is analyzed based on the operation of the system and considered in cost analysis in \cite{richard2018fast}, however, the battery degradation is not directly accounted for in the dispatch scheduling strategies. To solve this, Schneider $et\ al.$\cite{schneider2020rechargeable} proposed to implement the rainflow counting algorithm (RCA) in the scheduling model for dispatch optimization. However, due to the RCA having no analytical mathematical expression, the scheduling model doesn't explicitly include the battery degradation in the optimization problem but its effects are only analyzed post mortem. 

In this paper, we focus on minimizing the operational cost of FCS by combining the energy arbitrage model and the battery degradation model. In particular, we integrate the approximation of battery degradation based on state-of-charge (SoC) change as a penalty term directly in the optimization problem, thereby filling the current research gap. The FCS will charge the EVs and the BESS participates in the electricity market for energy arbitrage and it also mitigates the impulse of load demand. In this way, a shorter investment payback period and long-term profits of the FCS can be realized.

\section{Modeling approach}
In this section, we model the operation of battery storage systems in the FCS and degradation of batteries. As we try to investigate the impact of battery degradation on the optimal control of charging/discharging for the battery to minimize the overall operational cost of the system, we combine the energy arbitrage model with the battery degradation model. The demand profile is obtained from the predicted load demand in the FCS system and the electricity prices of Denmark are known a day in advance. Based on the load demand, the offering strategy will be determined, and the battery storage system will participate in the day-ahead energy market to buy energy in cheaper periods to keep it state-of-charge. More importantly, the energy arbitrage will also be considered if proved profitable to shift the energy consumption of the charging station away from the peaks of energy markets. Therefore, the dispatching problem in the paper is to determine the optimal scheduling strategy for the FCS based on the spot market prices and demand. Thus, in the scheduling model, the main part is the daily operation cost minimization based on the load profile and the operation of the battery storage system, the other part is the battery lifetime losses due to the operations. The two parts will be combined to build the final scheduling model to balance the energy arbitrage revenue, energy consumption cost and battery degradation cost. The details of the model are elaborated in the following sections.

\subsection{Battery operations in an FCS}
Considering the operations within a finite time horizon $t \in \mathcal{T} = \{1, 2,..., T\}$, the goal is to optimize the charge and discharge operations in the FCS in each time $t$ to minimize the overall operational cost for daily scheduling. The scheduling model for an FCS with a battery storage system within a day is presented in \mbox{(\ref{objetive})}. In the objective function of \mbox{(\ref{obj})}, the overall operation cost $J_t$ of the FCS system is composed of energy consumption cost with an additional degradation penalty term and the energy arbitrage revenue, where $P^{in}_t$ and $P^{out}_t$ represent the energy flow from and to the grid, $p_t$ is the electricity spot market price. It is minimized by determining the set of charging and discharging variables $P^{ch}_t$ and $P^{dis}_t$. The penalty term $\triangle c$ for the battery degradation is associated with specific degradation model, which will be elaborated in the following section. The scaling factor $a_k$ [DKK/kWh] is applied to the degradation penalty term for the operational cost of battery usage. The operation cost from energy consumption and energy arbitrage is modeled in the first term of the objective function. Constraint \mbox{(\ref{balance})} defines the energy flow from/to the grid based on the load demand and battery operations. The constraints for battery operation in the FCS system are presented by \mbox{(\ref{soc})}-\mbox{(\ref{pdis_m})}, where \mbox{(\ref{soc})} describes the evolution of battery SoC over the time horizon with to the variables $P^{ch}_t$ and $P^{dis}_t$, \mbox{(\ref{ini_end})} ensures the SoC at the end is equal to an expected value $SoC^{end}$, \mbox{(\ref{soc_lim})} limits the SoC within the battery capacity range, \mbox{(\ref{pch_m})} and \mbox{(\ref{pdis_m})} represent the power limits for charging and discharging.
\begin{mini!}|s|
{P^{ch}_t, P^{dis}_t}{\sum^T_{t=1}(P^{in}_t - P^{out}_t) \cdot p_t + a_k \cdot \triangle c}{\label{objetive}}{J_t = } {\label{obj}}{}
\addConstraint{P^{in}_t - P^{out}_t = P^{ch}_t - P^{dis}_t + P_{dem} \label{balance}}
\addConstraint{SOC_t = SOC_{t-1} + \frac{\tau}{C_{bat}}(\eta_{ch} P^{ch}_t - P^{dis}_t/\eta_{dis}) \label{soc}}
\addConstraint{SOC_{t=T} = SOC^{end} \label{ini_end}}
\addConstraint{0\leq SOC_t \leq 1 \label{soc_lim}}
\addConstraint{0\leq P^{ch}_t \leq P^{max} \label{pch_m}}
\addConstraint{0\leq P^{dis}_t \leq P^{max} \label{pdis_m}}.
\end{mini!}

\subsection{Battery degradation cost}
To account for the battery degradation in the energy arbitrage model, we utilized a DOD-based Peukert Lifetime Energy Throughput (PLET) model for estimating the battery capacity loss\cite{tran2013energy}. The PLET model involves the battery cycle with DOD based on Peukert's Law, where lifetime energy throughput is expressed as
\begin{equation}
\small
 C^{life}_{PLET} \triangleq n DOD^{k_p}
\label{Life_energy_throughput}
\end{equation}
where $n$ denotes the total number of cycles for charging and discharging within the lifetime of the battery, $k_p$ denotes the Peukert Lifetime constant between 1.1 and 1.3.

In the PLET battery degradation model, the total energy throughput $C^{life}_{PLET}$ obtained from experiment or datasheet is assumed to be constant for each specific DOD. Therefore, with the discretization of the equation \mbox{(\ref{Life_energy_throughput})}, the accumulated battery degradation based on PLET model for time period T could be estimated as
\begin{equation}
\small
 Q^{PLET}_{loss} = \frac{1}{C^{life}_{ij}} \sum^T_{t=1} (\triangle DOD(t))^{k_p}
\label{Loss}
\end{equation}
where $Q^{PLET}_{loss}$ represents the capacity loss, $\triangle DOD(t)$ denotes the SoC change at the time interval $t$.

Finally, by assigning the $Q^{PLET}_{loss}$ to the $\triangle c$ in \mbox{(\ref{objetive})}, the FCS scheduling model is built. 

\section{Case study}
\begin{figure}[H]
\centering
\subfigure[Load profile of FCS system and day-ahead spot market prices for a day.]{\includegraphics[width=0.32\linewidth]{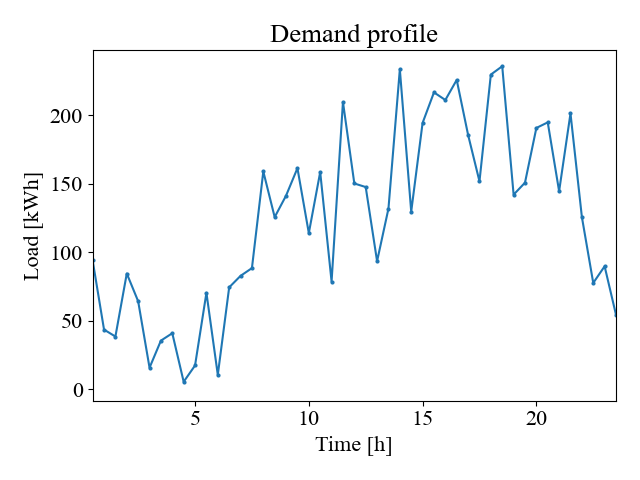}
\includegraphics[width=0.32\linewidth]{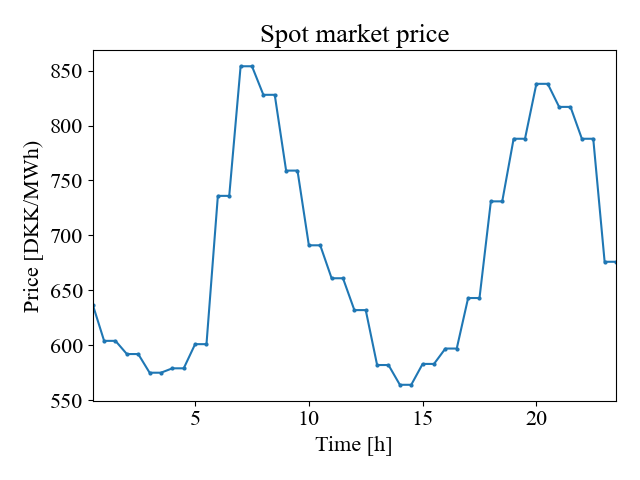}
\label{dem}}

\subfigure[Optimal dispatch schedule without battery usage penalty ($a_k=0$ DKK/kWh)]{\includegraphics[width=0.32\linewidth]{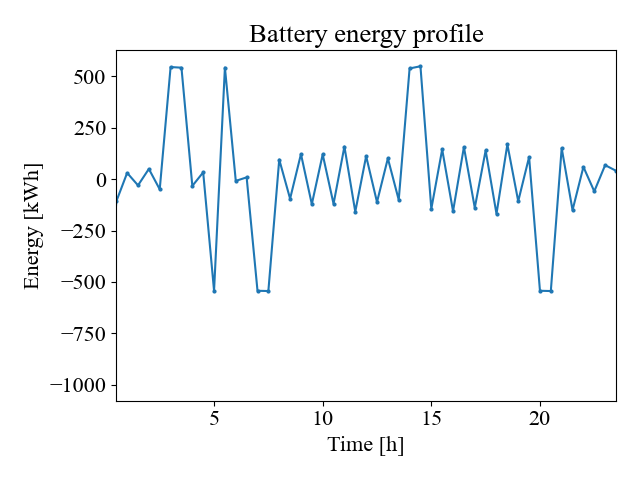}
\includegraphics[width=0.32\linewidth]{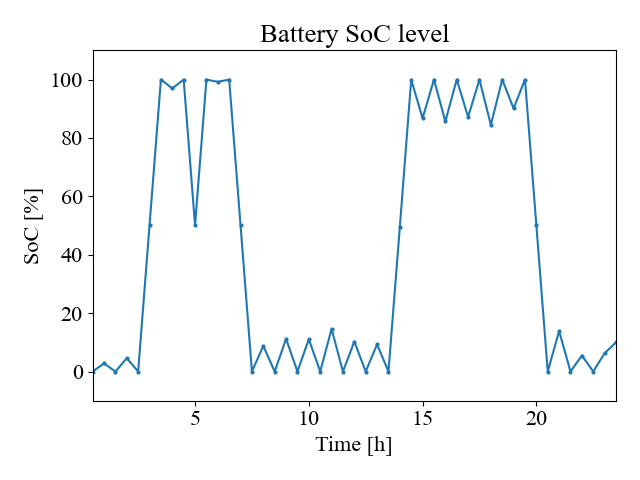}
\label{bat0}}

\subfigure[Optimal dispatch schedule when $a_k$=1 DKK/kWh ]{\includegraphics[width=0.32\linewidth]{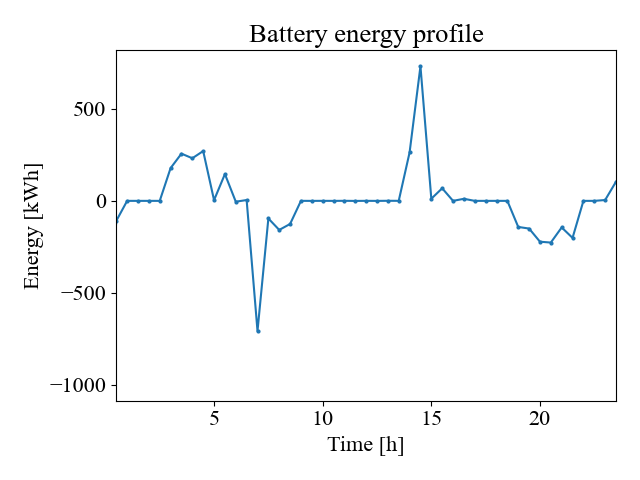}
\includegraphics[width=0.32\linewidth]{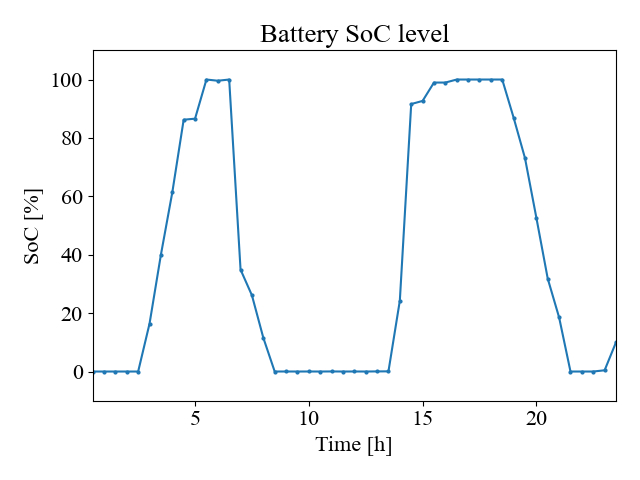}
\label{bat1}}

\subfigure[Optimal dispatch schedule when $a_k$=10 DKK/kWh ]{\includegraphics[width=0.32\linewidth]{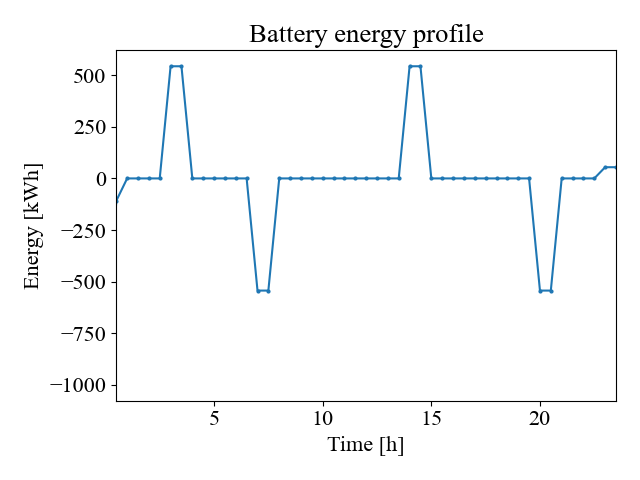}
\includegraphics[width=0.32\linewidth]{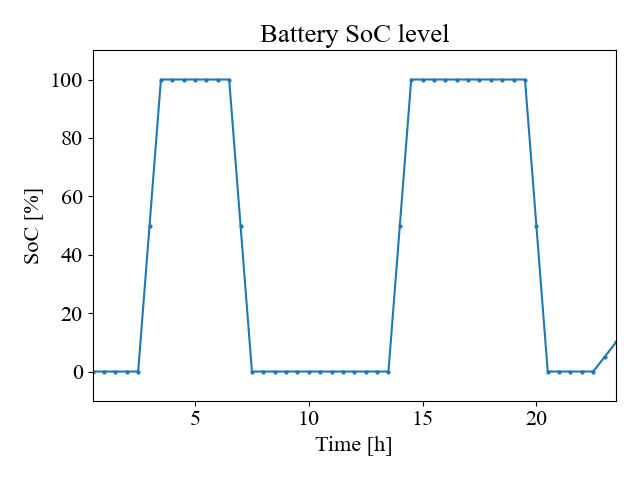}
\label{bat2}}
\caption{Optimal dispatch schedule for a given day with different values of penalty coefficient.}
\end{figure}

In the case study, the proposed method was verified and the results were obtained for an electricity service provider in Denmark using the predicted load data and day-ahead spot market prices for a day, shown in Fig. \ref{dem}. For the parameters of batteries utilized in the case study, the $C^{life}_{PLET}$ is 12500 and the $k_p = 1.15$. The maximum energy flow $P^{max} = 1$ MW. The time interval is 30 min in this study, therefore $T = 48$ and yields 96 decision variables. Problem \mbox{(\ref{objetive})} is implemented in Python with Pyomo as a modeling interfacing \cite{hart2011pyomo}, and solved by IPOPT with MUMPS linear solver \cite{biegler2009large}. The problem is solved on a rolling horizon basis, i.e. every time step a new solution is derived.

As we mentioned, there is an additional economic cost for battery usage, corresponding to the impacts of operation in long-term degradation, which is assigned by a scaling factor $a_k$ [DKK/kWh]. To investigate the influence of battery usage on the dispatch scheduling strategy, different empirical values of the coefficient $a_k$ are assigned. The optimal value can be further determined by comparing the optimization results with a finite set of values. As is shown in Fig. \ref{bat0}, when $a_k = 0$ [DKK/kWh], which indicates the usage of battery is not accounted for in the operational planning. Thus, to minimize the overall operational cost of FCS, based on the load demand and electricity prices, the battery storage system participates fully in energy arbitrage by purchasing and storing more energy during lower price periods and shifting the bulky energy consumption of FCS away from peak prices. The total energy arbitrage revenue is 1233.9 DKK. In addition, due to the oscillation of load profile, the battery storage system will charge and discharge repeatedly to ensure energy balance. However, such operations will accelerate the degradation of the battery due to the cycle aging. Therefore, by assigning a value to the coefficient $a_k$ of the degradation penalty term, which is based on the SoC change, the optimal scheduling strategies will be influenced.

 The coefficient is then assigned with an empirical value $a_k = 1$ [DKK/kWh]. The resulting battery operations are shown in Fig. \ref{bat1}. It can be observed from the battery energy profile that the battery storage system still participates in energy arbitrage and helps reducing the consumption cost of FCS during peak prices period, while the repeating charging and discharging operations are reduced during the load oscillation period, thus SoC of the battery becomes stable and the lifetime loss of battery is reduced. In addition, the energy arbitrage revenue is reduced to 1196.8 DKK. It indicates the direct energy purchase from the grid for the FCS consumption when the spot market prices are relatively low is preferred due to the penalty of battery usage. Moreover, as is shown in Fig. \ref{bat2}, when the penalty coefficient $a_k$ increases to 10 [DKK/kWh], the battery usage is reduced further and the optimal dispatch schedule is significantly different. In this case, battery charging and discharging would be most desirable at 3 am, 2 pm when the spot market prices are lowest and at 7 am, 8 pm when the spot market prices are highest. And the energy arbitrage revenue is 1233.8 DKK which is almost the same as that without battery degradation. In other words, the battery degradation is reduced at the expense of more impacts of load fluctuations on the grid.

\section{Conclusion}
In this paper, an operation scheduling model for an FCS with battery storage system incorporating battery degradation is proposed. The model optimizes the battery dispatch schedules for energy arbitrage and shifts the energy consumption away from the peak prices, thus minimizing the overall operational cost while considering battery degradation. This is realized by including the degradation model based on the SoC change into the scheduling model. By assigning different values of coefficient for the penalty of battery usage, the scheduling strategies will be influenced. The case study demonstrated that adding a degradation penalty in the objective function of the scheduling model leads to a trade-off among battery degradation, energy arbitrage and consumption cost. This provides the foundation for the assessment of cost-effectiveness of battery storage system investment in FCS for long-term operation.

\bibliographystyle{IEEEtran}
\bibliography{IEEEabrv,Bibliography}

\begin{thebibliography}{10}
\providecommand{\url}[1]{#1}
\csname url@samestyle\endcsname
\providecommand{\newblock}{\relax}
\providecommand{\bibinfo}[2]{#2}
\providecommand{\BIBentrySTDinterwordspacing}{\spaceskip=0pt\relax}
\providecommand{\BIBentryALTinterwordstretchfactor}{4}
\providecommand{\BIBentryALTinterwordspacing}{\spaceskip=\fontdimen2\font plus
\BIBentryALTinterwordstretchfactor\fontdimen3\font minus
  \fontdimen4\font\relax}
\providecommand{\BIBforeignlanguage}[2]{{%
\expandafter\ifx\csname l@#1\endcsname\relax
\typeout{** WARNING: IEEEtran.bst: No hyphenation pattern has been}%
\typeout{** loaded for the language `#1'. Using the pattern for}%
\typeout{** the default language instead.}%
\else
\language=\csname l@#1\endcsname
\fi
#2}}
\providecommand{\BIBdecl}{\relax}
\BIBdecl

\bibitem{outlook2021accelerating}
I.~G.~E. Outlook, ``Accelerating ambitions despite the pandemic,''
  \emph{International Energy Agency: Paris, France}, 2021.

\bibitem{morrissey2016future}
P.~Morrissey, P.~Weldon, and M.~O’Mahony, ``Future standard and fast charging
  infrastructure planning: An analysis of electric vehicle charging
  behaviour,'' \emph{Energy Policy}, vol.~89, pp. 257--270, 2016.

\bibitem{mahfouz2019grid}
M.~M. Mahfouz and M.~R. Iravani, ``Grid-integration of battery-enabled dc fast
  charging station for electric vehicles,'' \emph{IEEE Transactions on Energy
  Conversion}, vol.~35, no.~1, pp. 375--385, 2019.

\bibitem{sbordone2015ev}
D.~Sbordone, I.~Bertini, B.~Di~Pietra, M.~C. Falvo, A.~Genovese, and
  L.~Martirano, ``Ev fast charging stations and energy storage technologies: A
  real implementation in the smart micro grid paradigm,'' \emph{Electric Power
  Systems Research}, vol. 120, pp. 96--108, 2015.

\bibitem{han2018economic}
X.~Han, Y.~Liang, Y.~Ai, and J.~Li, ``Economic evaluation of a pv combined
  energy storage charging station based on cost estimation of second-use
  batteries,'' \emph{Energy}, vol. 165, pp. 326--339, 2018.

\bibitem{ding2015value}
H.~Ding, Z.~Hu, and Y.~Song, ``Value of the energy storage system in an
  electric bus fast charging station,'' \emph{Applied Energy}, vol. 157, pp.
  630--639, 2015.

\bibitem{chen2018coordinated}
H.~Chen, Z.~Hu, H.~Zhang, and H.~Luo, ``Coordinated charging and discharging
  strategies for plug-in electric bus fast charging station with energy storage
  system,'' \emph{IET Generation, Transmission \& Distribution}, vol.~12,
  no.~9, pp. 2019--2028, 2018.

\bibitem{richard2018fast}
L.~Richard and M.~Petit, ``Fast charging station with battery storage system
  for ev: Grid services and battery degradation,'' in \emph{2018 IEEE
  International Energy Conference (ENERGYCON)}.\hskip 1em plus 0.5em minus
  0.4em\relax IEEE, 2018, pp. 1--6.

\bibitem{schneider2020rechargeable}
S.~F. Schneider, P.~Nov{\'a}k, and T.~Kober, ``Rechargeable batteries for
  simultaneous demand peak shaving and price arbitrage business,'' \emph{IEEE
  Transactions on Sustainable Energy}, vol.~12, no.~1, pp. 148--157, 2020.

\bibitem{tran2013energy}
D.~Tran and A.~M. Khambadkone, ``Energy management for lifetime extension of
  energy storage system in micro-grid applications,'' \emph{IEEE Transactions
  on Smart Grid}, vol.~4, no.~3, pp. 1289--1296, 2013.

\bibitem{hart2011pyomo}
W.~E. Hart, J.-P. Watson, and D.~L. Woodruff, ``Pyomo: modeling and solving
  mathematical programs in python,'' \emph{Mathematical Programming
  Computation}, vol.~3, no.~3, pp. 219--260, 2011.

\bibitem{biegler2009large}
L.~T. Biegler and V.~M. Zavala, ``Large-scale nonlinear programming using
  ipopt: An integrating framework for enterprise-wide dynamic optimization,''
  \emph{Computers \& Chemical Engineering}, vol.~33, no.~3, pp. 575--582, 2009.

\end{thebibliography}

\end{document}